\documentclass[groupedaddress,
 aip,
 rsi,
 amsmath,amssymb,
 reprint,%
]{revtex4-1}

\usepackage{graphicx}% Include figure files
\usepackage{dcolumn}% Align table columns on decimal point
\usepackage{bm}% bold math
\usepackage[utf8]{inputenc}
\usepackage[T1]{fontenc}
\usepackage{mathptmx}
\usepackage{etoolbox}
\usepackage{amsmath}

\graphicspath{{Figs/}}
\usepackage{float}
\usepackage[utf8]{inputenc}
\usepackage{color}
\usepackage{nicefrac}

\makeatletter
\def\@email#1#2{%
 \endgroup
 \patchcmd{\titleblock@produce}
  {\frontmatter@RRAPformat}
  {\frontmatter@RRAPformat{\produce@RRAP{*#1\href{mailto:#2}{#2}}}\frontmatter@RRAPformat}
  {}{}
}%
\makeatother
\begin{document}

%\preprint{AIP/123-QED}

\title{Laser-induced fluorescence detection of the  4d$^4D_{7/2}\rightarrow$5p$^4P^\circ_{5/2}$ transition of Kr II in a hollow cathode discharge}

\author{Y. Dancheva}
\email{The author to whom correspondence may be addressed: yordanka.dancheva@aerospazio.com}
%\affiliation{DSFTA, University of Siena, via Roma 56, Siena, Italy}
 
\author{P. Coniglio}
%\affiliation{Aerospazio Tecnologie S.r.l. - Rapolano Terme, Italy}
 
\author{D. Pagano}
%\affiliation{Aerospazio Tecnologie S.r.l. - Rapolano Terme, Italy}

\author{A. Garde}
%\affiliation{Aerospazio Tecnologie S.r.l. - Rapolano Terme, Italy}

\author{F. Scortecci}
%\affiliation{Aerospazio Tecnologie S.r.l. - Rapolano Terme, Italy}

\affiliation {Aerospazio Tecnologie Srl, via dei Tessili 99, 53040  Rapolano Terme (SI), Italy}

\date{\today}% It is always \today, today,

\begin{abstract}
Electric propulsion requires exhaustive ground test campaigns to obtain an accurate characterization of the propulsion devices, or thrusters, used by the spacecraft. Among the many plasma parameters, accurately measured during the tests, that of the ion velocity is key, and can be measured using non-intrusive tools such as Laser-Induced Fluorescence (LIF) diagnostics. The ion velocity is inferred by  Doppler shift measurements that presupposes a precise and accurate knowledge of the wavelength of excitation of the ions at rest. Today electric propulsion is moving towards the use of Krypton as a propellant, due to the dramatic increase in cost of the more advantageous Xenon gas propellant, commonly used until now. In this work the 4d$^4D_{7/2}\rightarrow$5p$^4P^\circ_{5/2}$ transition of Kr II is used for LIF diagnostic in hollow cathode discharge.

\end{abstract}
\smallskip

\maketitle

\section{Introduction}
Traditionally, Xenon has been the favorite propellant gas for electric propulsion applications as it provides an optimal compromise between performance and ease of handling. Although Xenon has numerous technical advantages, its relative scarcity and resulting increase in price are now posing considerable budgetary constraints for its use, especially for deep space exploration \cite{Tirila_2023, Nakles_2011}. Therefore, a number of alternative propellants are being examined intensively for use in electrostatic spacecraft propulsion thrusters \cite{Duchemin_2009}. A more economic alternative to Xe needs to be identified, while maintaining thruster performance levels, and without overlooking performance optimization aspects. Among various types of electric
thrusters, the Hall Effect Thruster (HET) is one of the most widely used
owing to its simple structure, high reliability, and long life \cite{Mazouffre_2016, Levchenko_2018}.

Like Xenon, Krypton is a noble gas and can be integrated easily into existing spacecraft propellant management systems with minimal modifications. As this gas has similar ionization potential it should not dramatically affect thruster efficiency, and the lower atomic mass could possibly produce a 25\% increase in specific impulse due to the increased propellant exit velocity of lighter ions \cite{Linnell_2007, Hargus_2012}. The higher specific impulse also provides advantages in space missions such as maintaining spacecraft orbit \cite{Kurzyna_2018, Lim2019}. A lower anode efficiency of approximately 5-15$\%$ is expected with respect to Xenon propellant at the same operating conditions \cite{Hargus_2012, Kurzyna_2018, Jorns_2022}. Krypton is also about ten times more common in the atmosphere than Xenon, and hence  less expensive. Currently, Krypton is the most widely-used alternative propellant in HET both for space applications (for example the Starlink constellation), as well as ground testing \cite{Tirila_2023}. Table$\nobreakspace$\ref{Xe Kr} summarizes the properties of Xenon and Krypton specifically relevant to electrostatic spacecraft propulsion. 
\begin{table}[h]
\centering
 \caption{\label{Xe Kr}Kr and Xe properties comparison}
    \begin{tabular}{|l|c|c|}
    \hline
    \hline
    & & \\
     \textbf{Property}  & \textbf{Xe}  & \textbf{Kr}\\
      & & \\
      \hline 
      \hline
      Atomic mass & 131.293$\nobreakspace$amu  &  83.798$\nobreakspace$amu\\
      \hline
    First ionization energy & 12.1$\nobreakspace$eV  &  14.0$\nobreakspace$eV\\
      \hline
       Atmospheric & & \\
       
       concentration & 87$\nobreakspace$ppb  &  1000$\nobreakspace$ppb\\
       \hline
        Stable isotopes & 9  &  6\\
        \hline
        Odd isotopes & 2 & 1 \\
        \hline
        Critical pressure  & 57.65$\nobreakspace$atm & 54.3$\nobreakspace$atm\\
            \hline
            Critical temperature & 290$\nobreakspace$K & 209$\nobreakspace$K\\
            \hline
            Boiling point (1$\nobreakspace$atm) & 161$\nobreakspace$K & 120$\nobreakspace$K\\
            \hline
          \end{tabular}
 \end{table}

 The characterization and qualification of ion thrusters require extensive, long-duration test campaigns in space simulators. A valuable tool for characterizing thruster performance and comparing this with already-developed models is Light-Induced Fluorescence (LIF) spectroscopy, which measures the velocity of ions employing the Doppler effect. Many excellent works on LIF characterization of ion thrusters are available in the literature, but only few of them consider thrusters operating with Kr propellant. Advantageous transitions for LIF diagnostics are characterized by large intensity and narrow intrinsic width (transitions involving long-living levels). The Xe ion transition at 834$\nobreakspace$nm (5d$^2[4]_{7/2}\rightarrow$6p$^2[3]^{\circ}_{5/2}$) is used for LIF diagnostics by the majority of the electric propulsion research community. As to Kr ion, such favoured transitions are at 820$\nobreakspace$nm (the 4d$^4F_{7/2}\rightarrow$5p$^2D^\circ_{5/2}$ transition) \cite{Lejeune_2012} and at 729$\nobreakspace$nm (the 4d$^4D_{7/2}\rightarrow$5p$^4P^\circ_{5/2}$ transition) \cite{Hargus_2011, Buorgeois_2011}. An accurate knowledge of the features of the transition used for LIF spectroscopy is required for better determination of the ion velocity. 
 The first task to solve is to find a suitable source of Kr ions with null drift velocity. One possible Kr$^+$ source is a hollow cathode plasma source, as the one used in this work, operated normally as a neutralizer in the electric propulsion.
 
  LIF diagnostic has been previously applied for characterization of ions emitted from a hollow cathode in the plume region (see for example \cite{Potrivitu_2019, Dodson_2018} and reference therein). It has been shown \cite{Potrivitu_2019} that when operating in the so-called diode configuration (with an external anode that closes the electrical circuit) the ion velocity, measured starting from the exit plane of the cathode to the near field, can be modified. Axial velocity components with opposite signs have been measured. It has been shown that the anode acts as a physical boundary for the cathode plasma discharge and drives the discharge current balance through its collection surface. The electric field topology in consequence can be influenced in order to fulfill stable discharge condition. Moreover, the shape of the anode modifies the "view factor" from the cathode orifice, which influences the electron and ion fluxes, and hence the plasma properties in the cathode-anode gap. Similar results have been observed in the exit plane of hollow cathode assemblies developed at the University of Michigan \cite{Dodson_2018}, where it has been shown that the presence of high-energy ions is consistent with a potential-hill model. It should be pointed out that the axial velocity drift value strongly depends on many parameters among which the hollow cathode design and operational power, the anode shape and the distance to it, the cathode regime of operation, etc.
  
 The air wavelength of the  Kr II 4d$^4D_{7/2}\rightarrow$5p$^4P^\circ_{5/2}$ transition has been previously measured to be 728.982$\nobreakspace$nm with an accuracy of 0.7$\nobreakspace$pm (about 400$\nobreakspace$MHz) \cite{Dzierżęga_2001, Saloman_2007}. Such accuracy would introduce an error of about 300$\nobreakspace$m/s in the determination of the Kr ion velocity.
 In this work an attempt is performed to better determine the  vacuum/air wavelength of the Kr II transition 5d$^4D_{7/2}\rightarrow$5p$^4P^\circ_{5/2}$. For this purpose, the LIF  spectroscopy is applied a hollow cathode discharge, namely behind the keeper plate in the plasma region next to the cathode orifice. Further investigation is necessary to verify whether the region selected is free of ion velocity drift.

\section{Set-up}
The LIF is a particularly powerful optical-diagnostic tool providing valuable, species selective information with excellent spatial resolution. A highly coherent, single-frequency, laser source that can be tuned over a broad spectral interval permits to establish a versatile laser-plasma interaction tool. Given its remote detection nature, LIF diagnostic is minimally invasive. The plasma is interrogated locally with an electromagnetic tool: all the optical devices used to illuminate the plasma and to detect its fluorescence are placed far from the interaction point,  can have dimensions in the mm scale, and the interaction volume can have sizes as small as few millimeters. The ion velocity distribution function (IVDF) can be measured along different directions with excellent spatial and time resolution. The measurement procedure is based on detecting the intensity of the fluorescence emitted by the plasma in the interaction point, in response to the illumination by means of narrow-band radiation characterized by an accurately and precisely known frequency that can be scanned within an opportune interval.

\subsection{\label{test facility}The test facility}
\begin{figure}[h]
\centering
\includegraphics[width=0.5\textwidth]{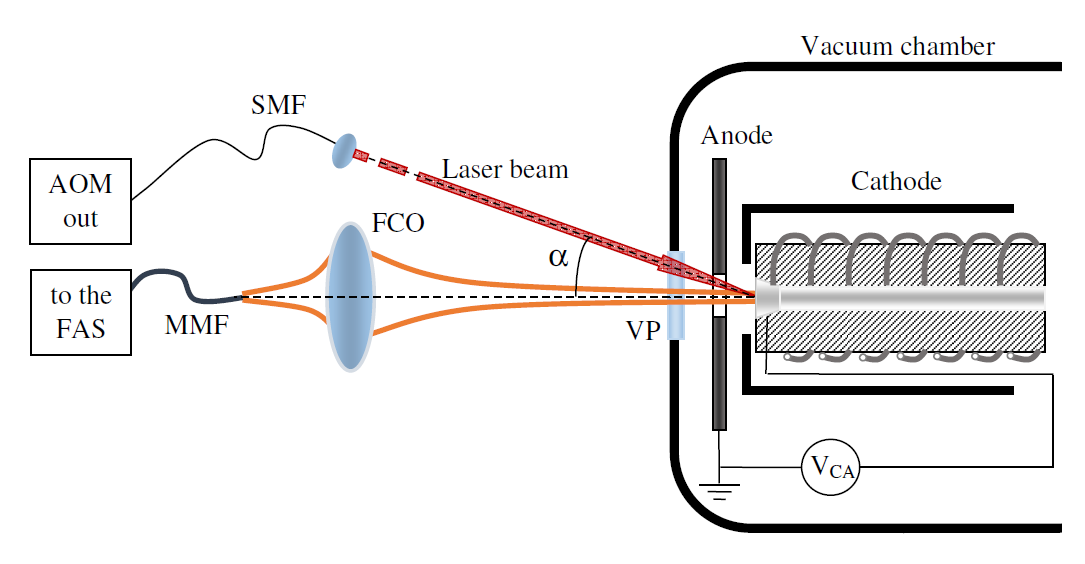}
\caption{Schematics of the laser excitation and fluorescence collection arrangement. FAS - fluorescence analyzing system, FCO - fluorescence collection objective, VP - view port of the vacuum chamber, SMF, MMF - single-mode and multi-mode fibers and AOM - acousto-optical modulator.}
\label{set-up}
\end{figure}

The measurements shown in this work are conducted in a non-magnetic stainless steel vacuum chamber. Vacuum is obtained by a single stage cryogenic panel. The base pressure of the vacuum chamber is as low as $10^{-7}$mbar and increases up to about $10^{-5}$mbar during the cathode operation. The hollow cathode is operated in diode regime (see Fig.\ref{set-up}) at currents of about 3$\nobreakspace$A with an anode positioned  20$\nobreakspace$mm away from it. The measurements are performed positioning the LIF equipment outside the vacuum chamber illuminating the detection region and detecting the emitted fluorescence through an optical access (view port).

\subsection{The LIF set-up}
\label{LIF set-up}
The light source used to excite the  4d$^4D_{7/2}\rightarrow$5p$^4P^\circ_{5/2}$ Kr$\nobreakspace$II transition is a tunable diode laser in a primary-and-secondary configuration (the primary laser is an extended-cavity diode laser and the secondary laser is a tapered amplifier) with a linewidth better than 1$\nobreakspace$MHz.

\begin{figure}[ht!]
\centering
\includegraphics[width=0.5\textwidth]{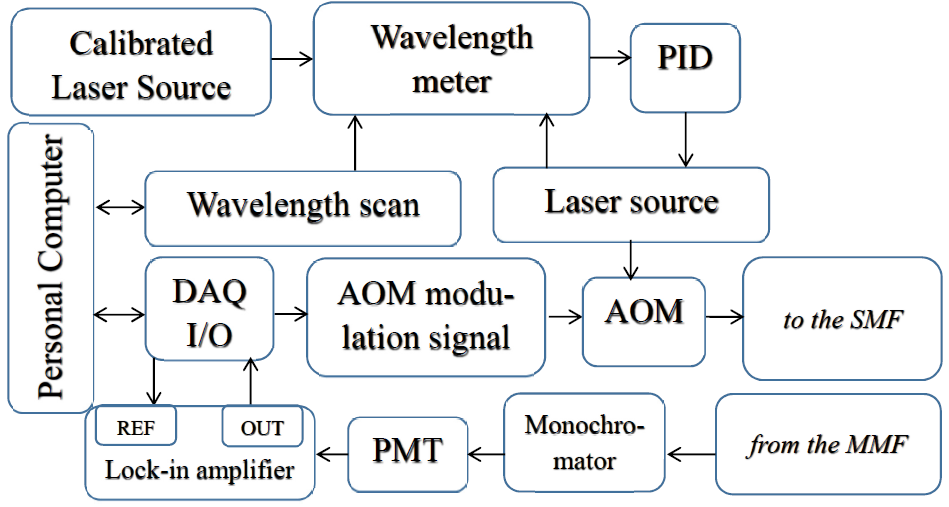}
\caption{Schematic of the LIF set-up: Proportional-Integral-Derivative (PID) controller; acousto-optical modulator (AOM); photo-multiplier tube (PMT); and data acquisition card (DAQ).}
\label{set-up laser}
%\vspace{-1em}
\end{figure}
A schematic of the LIF set-up is given in Fig.\ref{set-up laser} (a more detailed description can be found in Refs.\cite{Dancheva_2022, Dancheva_2023}). The laser wavelength is locked and scanned with the help of a high accuracy ($\pm\nobreakspace10\nobreakspace$MHz) wavelength meter and a proportional-integral-derivative (PID) controller.  The wavelength meter is calibrated periodically using a diode laser that is frequency stabilized to the  absorption profile of the Caesium D$_2$ line and typically drifts below 2$\nobreakspace$MHz/day.

The detection region is near to the cathode exit plane at the cathode tip center, where the laser beam and the fluorescence collection objective (FCO) view regions are crossing (see Fig.\ref{set-up}). The laser beam is aligned at a small angle $\alpha$ with respect to the normal to cathode exit plane. The laser intensity is amplitude modulated at 3$\nobreakspace$kHz by a fibre-coupled, acousto-optical modulator (AOM).  The AOM output is coupled to a single-mode optical fiber  (see Fig.\ref{set-up laser}) that provides a laser beam with waist diameter of 0.3$\nobreakspace$mm using a fiber collimator positioned at 0.6$\nobreakspace$m away from the cathode. 

The view spot of the FCO is about 2.5$\nobreakspace$mm in diameter at the region of detection. The fluorescence signal at air wavelength of 473.9$\nobreakspace$nm (corresponding to the transition 5p$^4P^\circ_{5/2}\rightarrow$5s$^4P_{5/2}$) is  selected using a grating monochromator. Subsequently, the signal at 473.9$\nobreakspace$nm is detected using a photo-multiplier tube (PMT). The LIF signal demodulation is performed by a lock-in amplifier, whose input is the  PMT current, converted to voltage by a trans-impedance amplifier. The poor signal-to-noise ratio of the LIF spectra is counteracted by using a long lock-in integration time, implying a slow scan of the laser wavelength and averaging over many measurements. Indeed, the lock-in amplifier integration time is 300$\nobreakspace$ms, which with a third order low-pass filter at the output determines a bandwidth of about 0.3$\nobreakspace$Hz .

The laser vacuum wavelength is locked and scanned by the wavelength meter using a PID control (the schematics of the set-up are shown in Fig.\ref{set-up laser}). A scan of 20$\nobreakspace$pm (about 11$\nobreakspace$GHz) is performed backwards and forwards using the wavelength meter in an overall time of 100$\nobreakspace$s. The bidirectional scanning permits to account for possible shifts of the spectral line due to integration time.

The wavelength meter used in this work is temperature and pressure stabilized. A variation less than 50$\nobreakspace$MHz is expected for temperature variations of about 10$^\circ$C, which are well compensated for rates lower than \nicefrac{2$^\circ$C}{hour} that is higher than the typical temperature excursion in the laboratory.

\section{The K\lowercase{r} II 4\lowercase{d}$^4$D$_{7/2}\rightarrow$5\lowercase{p}$^4$P$^\circ_{5/2}$ line.}
In this analysis the Kr isotopes with concentration higher than 0.3$\%$ are considered and listed in Table \ref{isotopes}. A theoretical line-shape is calculated and used to fit the experimental data.
\begin{table}[h]
    \centering
        \caption{Isotope shift and hyperfine coefficients}
    \begin{tabular}{|l|c|c|}
     \hline
      \hline
       &  \textbf{Relative shift $\delta_i$} & \\
      \textbf{Isotope} & with respect to & \textbf{Abundance $\alpha$}\\
      & $^{84}$Kr (MHz) & ($\%$)\\
      \hline
      78 &  1185.7 & 0.35\\
       \hline
       80 & 768.0 & 2.27\\
        \hline
       82 & 372.6 & 11.56\\
        \hline
       83 & 175.0 & 11.55 \\
        \hline
       84 &  0 & 56.9\\
        \hline
       86 &  -365.2 & 17.37\\
        \hline
         \hline
        \textbf{Electronic} &  \textbf{A coefficient} & \textbf{B coefficient}\\
       \textbf{state} & (MHz) & (MHz)\\
         \hline
         4d$^4$D$_{7/2}$ & -43.513 & -294.921 \\
      \hline
      5p$^4$P$_{5/2}$ & -167.2 & +91\\
       \hline
        \hline
   \end{tabular}
    \label{isotopes}
\end{table}

Only one isotope ($^{83}$Kr) has a non zero nuclear spin ({\bf I}=9/2) and thus its hyperfine structure should be included in the line-shape analysis. The hyperfine constants of both the states of the $^{83}$Kr II transition have been measured previously \cite{Scholl_1986, Schuessler_1992} and are given in Table \ref{isotopes}.
\begin{table}[h]
\centering
 \caption{\label{HFS}Kr hyperfine structure components}
    \begin{tabular}{|c|c|c|c|c|}
    \hline
    \hline
       &   & \textbf{Relative} & \textbf{Relative} & \\
      &  & \textbf{shift} & \textbf{shift $\delta_i$} & \textbf{Intensity}
     \\
     \textbf{F} & \textbf{F$^\prime$} & \footnotesize{(with respect}   & \footnotesize{(with respect} & \textbf{ $\beta$}($\%$) \\
     & &  \footnotesize{to J$\rightarrow$J$^\prime$)} (MHz) &   \footnotesize{to $^{84}$Kr)} (MHz) &  \\
      \hline 
      \hline
      8 & 7 & -1099.2 & -924.2 & 0.21 \\
      \hline
      7 & 7 & -1559.6 & -1384.6 & 0.034 \\
      \hline
      7 & 6 & -431.7 & -256.7 & 0.15 \\
      \hline
      6 & 7 & -1901.1 & -1726.1 & 0.0031 \\
      \hline
      6 & 6 & -773.2 & -598.2 & 0.055 \\
      \hline
      6 & 5 & 223.2 & 398.2 & 0.10 \\
      \hline
      5 & 6 & -1020.2 & -845.2 & 0.0083 \\
      \hline
      5 & 5 & -23.8  & 151.2 & 0.064 \\
       \hline
      5 & 4 & 827.3 & 1002.3 & 0.065 \\
       \hline
      4 & 5 & -197.5 & -22.5 & 0.015 \\
       \hline
      4 & 4 & 653.7 & 828.7 & 0.063 \\
       \hline
      3 & 4 & 535.8 & 710.8 & 0.022 \\
       \hline
      3 & 3 & 1230.4 & 1405.4 & 0.052 \\
       \hline
      3 & 2 & 1759.3 & 1934.3 & 0.013 \\
       \hline
      2 & 3 & 1154.3 & 1329.3 & 0.03 \\
       \hline
      2 & 2 & 1683.2 & 1858.2 & 0.033 \\
       \hline
      1 & 2 & 1638.3 & 1813.3 & 0.038 \\
      \hline
       \hline
          \end{tabular}
          \label{table HFS}
 \end{table}

\begin{figure}[h]
\centering
\includegraphics[width=0.5\textwidth]{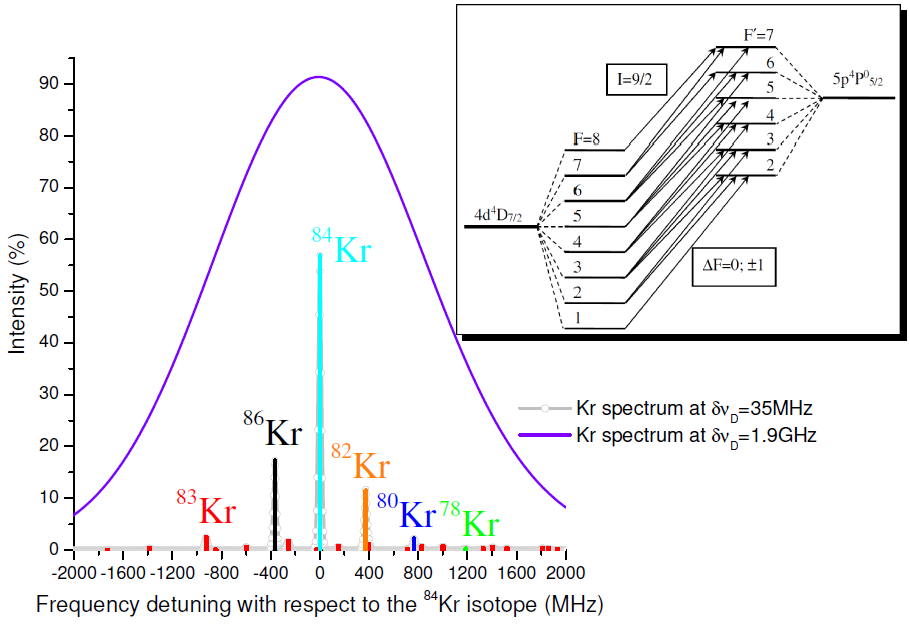}\\
\caption{The transition line-shape at two different temperature/Doppler widths. The hyperfine structure of $^{83}$Kr is shown in the inset.}
\label{line}
\end{figure}

As can be seen from Fig.\ref{line} the lower $^{83}$Kr II state splits in 8 hyperfine sublevels and the upper one in 6 with total angular momenta as denoted on the energy schematic. The electric-dipole allowed 18 transitions are shown in the plot. A detailed description about the hyperfine structure and the relevant selection rules can be found elsewhere \cite{Arimondo_1977, Hargus_2010, Hargus_2001}. The spectrum of the 4d$^4D_{7/2}\rightarrow$5p$^4P^\circ_{5/2}$ transition at Doppler linewidth of 35$\nobreakspace$MHz (to highlight the line structure) and 1900$\nobreakspace$MHz (to show the line-shape at ion temperatures near to the measured one) are reported in Fig.\ref{line}.

\section{Results}
\begin{figure}[h!]
\centering
\includegraphics[width=0.5\textwidth]{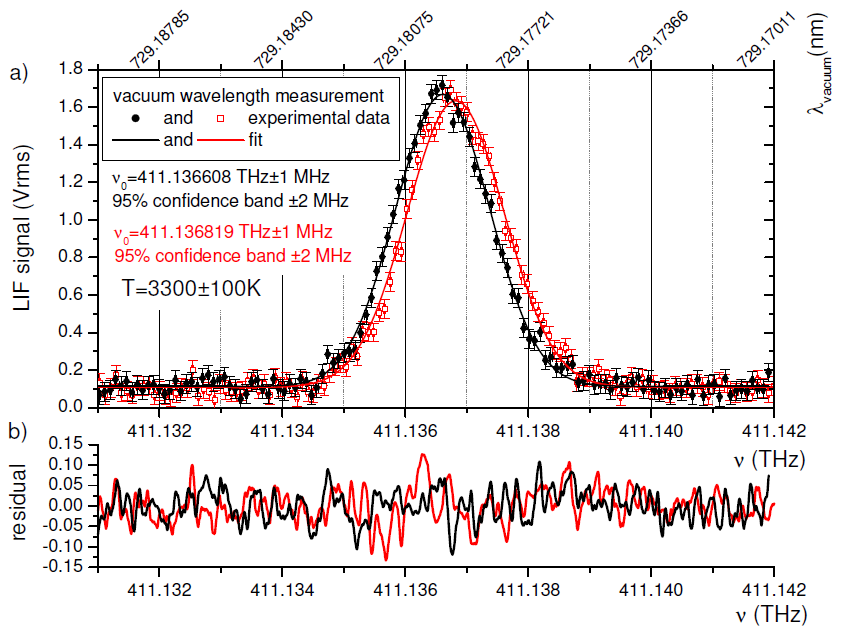}\\
\includegraphics[width=0.5\textwidth]{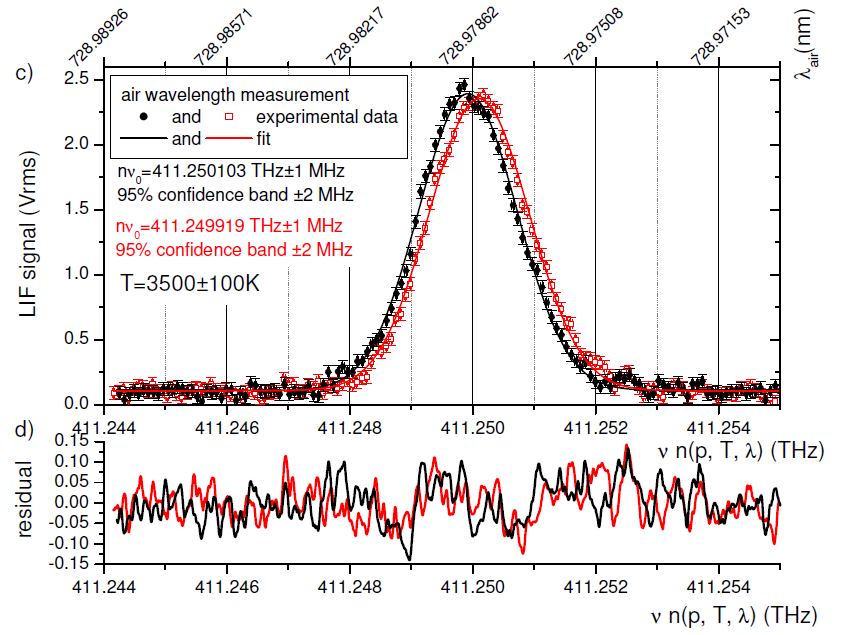}
\caption{Kr II line at about 729$\nobreakspace$nm when scanning the laser wavelength in both directions (upwards ramp - black traces, downward ramp - red traces) and measuring the vacuum a) and the air wavelength c). For the sake of convenience, the measurements are also presented as a function of the frequency - vacuum case, and as a function of $\nu n(p,T,\lambda)$ - air case, where {\it n} is the index of refraction of the dry air at the wavelength of interest.  Each measurement is an average over 36 laser scans. The residual is shown for each case in plots b) and d).}
\label{measured_freq}
\end{figure}

 Fig.\ref{measured_freq} shows the spectrum of the 4d$^4D_{7/2}\rightarrow$5p$^4P^\circ_{5/2}$ recorded when scanning the laser frequency in both directions. Each trace is obtained as an average over 36 laser wavelength scans and the cathode is kept at constant operational parameters during all the measurements.
The spectrum is registered measuring both the vacuum and the air wavelength (at standard dry air: T=15$^\circ$C and p=760$\nobreakspace$mmHg). The experimental data are fitted using an orthogonal distance regression algorithm \cite{Pallavi_2022} considering the wavelength meter accuracy and the LIF signal noise outside the spectral line as errors. 

The fitting function is given by \cite{Demtroeder_2013}:
\begin{equation}
\begin{split}
   G(\nu)=A\sum_{i}B_i\exp^{-4\ln{2}\big(\frac{\nu-(\nu_{0}\pm\delta_i)}{\Delta\nu_D}\big)^2}+D,\\
   \nobreakspace \Delta\nu_D=\frac{\nu_0}{c}\sqrt{8\ln{2}\frac{k_BT}{m}} ,
    \end{split}
\end{equation}
where the $\delta_i$ is the relative shift for each spectral component (given in Table \ref{isotopes} and \ref{table HFS}), $B_i$  are the relative intensities ($B_i$=$\alpha_i\times\beta_i$), $k_B$ is the Boltzmann constant, $T$ is the ion temperature, $m$ is the ion mass, $A$ is a scaling factor, $c$ is the speed of light, and $D$ accounts for possible data offset (see Fig.\ref{measured_freq}). Among the four fitting parameters ($A$, $\nu_0$, $T$, and $D$) the line central frequency $\nu_0$ and the ion temperature $T$ are of interest. The residuals, given in Fig.\ref{measured_freq}, are indicative for the fit goodness.

Figure \ref{measured_freq} shows the measured $\nu_0$ when scanning the laser wavelength in both directions. A shift, occurring due to the finite lock-in settling time of about 100$\nobreakspace$MHz is estimated from the fitted $\nu_0$ values. As this instrumental shift occurs symmetrically in both cases, the exact value of the $\nu_0$ has to be estimated as the average of the two results.

\begin{figure}[h]
\centering
\includegraphics[width=0.5\textwidth]{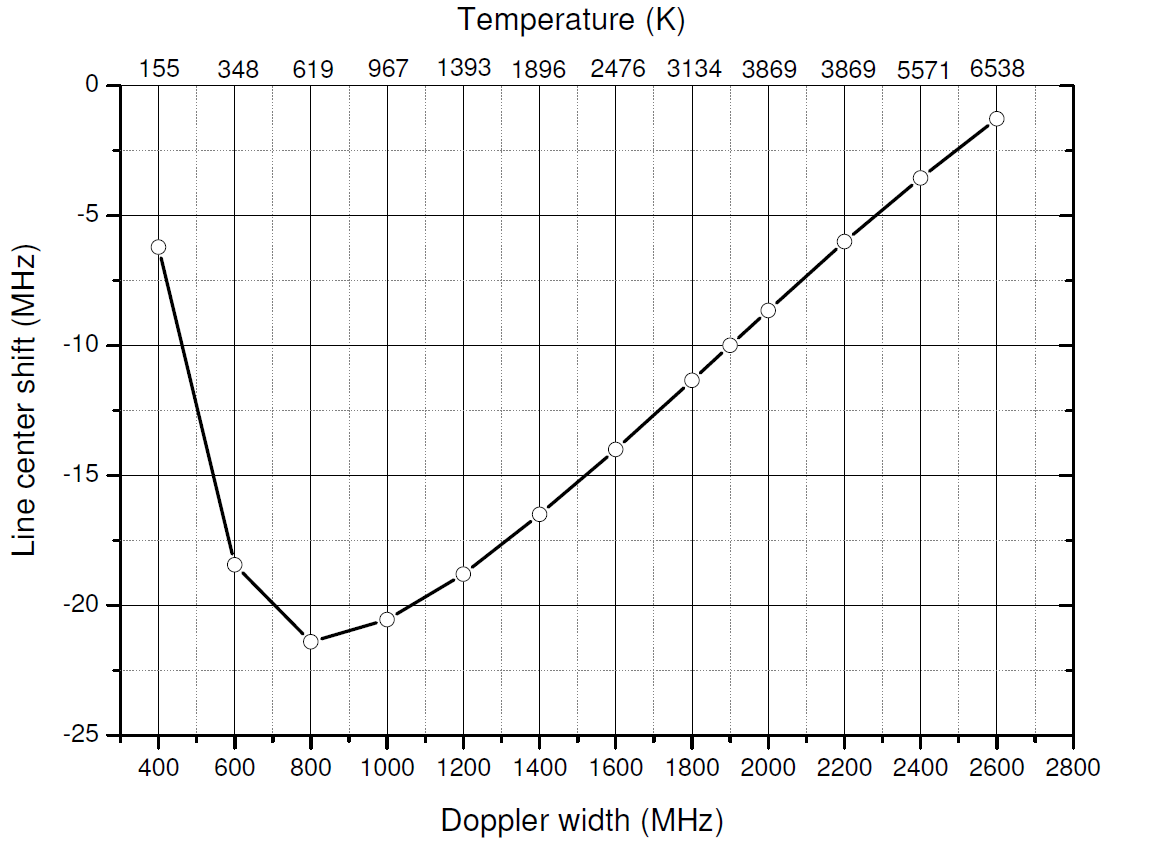}
\caption{Kr line center shift, as a function of the ion temperature (Doppler width).}
\label{shift HFS}
\end{figure}

The $\nu_0$ value is temperature dependent through the hyperfine and isotope structure contribution. A negative systematic error in the determination of $\nu_0$ is introduced at different Doppler widths (ion temperatures). The calculated shift is given in Fig.\ref{shift HFS}.

Considering a coverage factor of 2, for a confidence level of approximately 95$\%$, the measured central wavelength $\lambda_0$ and frequency $\nu_0$ are given in Table \ref{lambda}. Taking into account a Doppler width of 1900$\nobreakspace$MHz the corrected $\lambda_0$ values are also provided in Table \ref{lambda}.

\begin{table}[h]
    \centering
    \caption{Measured central wavelength and frequency. HFS - hyperfine structure.}
    \begin{tabular}{|l|c|c|}
    \hline
    \hline
    \bf{Condition:}& \bf{$\lambda_0$ (nm)} & \bf{Uncertainty (fm)} \\
    \hline 
    \hline 
      vacuum   & 729.179487 & 34  \\
      \hline
      air & 728.978602 & 34\\
      \hline
      \hline
       \bf{Condition:}& \bf{$\nu_0$(THz)} & \bf{Uncertainty (MHz)}\\
      \hline \hline
      vacuum/air & 411.136714 & 20 \\
        \hline\hline
         \bf{Condition:}& \bf{$\lambda_0$(nm)}& \bf{Uncertainty(fm)}\\
        \hline \hline
        vacuum \footnotesize{(corrected for} & &\\
        \footnotesize{HFS and isotope shift)} & 729.179504 & 34\\
                \hline
        air \footnotesize{(corrected for} & &\\
         \footnotesize{HFS and isotope shifts)} & 728.978619 & 34\\
               \hline \hline
       \end{tabular}
    \label{lambda}
\end{table}

The fitted ion temperature results in the range from 3300$\nobreakspace$K to 3500$\nobreakspace$K and can be related to the cathode wall temperature, which is about 1.5 times lower.

\section{Conclusions}
LIF diagnostic is performed in the plasma of a hollow cathode using the the Kr II 4d$^4D_{7/2}\rightarrow$5p$^4P^\circ_{5/2}$ transition. By performing long-lasting measurements an improved signal to noise is obtained as to make the residual noise contribute negligibly to the error of the measurements. The theoretical spectrum of the transition is calculated and used to fit the experimental data. Both the frequency and the vacuum/air wavelength centers are given. Further investigation is necessary to verify whether a null drift velocity can be expected in the orifice region of a hollow cathode operating in diode mode. In fact, work is in progress to evaluate a possible non-zero average Kr ion velocity by applying a counter-propagating laser excitation scheme in a proper Kr ion source.

The systematic error due to the presence of hyperfine and isotope structure in determination of the line centre is calculated and taken into account as well as its variation with the ion temperature.

\section{Acknowledgements}
This work has received funding from the European Union’s Horizon 2020 research and innovation programme under grant agreement No 101004140. The authors would like to thank Patricia Nugent for revising the English of the manuscript.

\section*{References}

\bibliography{refs}% Produces the bibliography via BibTeX.
\bibliographystyle{ieeetr}

\end{document}